\documentclass[letterpaper, 10 pt, conference]{ieeeconf}  

\IEEEoverridecommandlockouts                              

\usepackage{amsmath} 
\usepackage{amssymb}  
\usepackage{bm}
\usepackage{units}
\usepackage{hyphsubst}
\usepackage{nicefrac}

\usepackage{graphics} 
\usepackage[pdftex]{graphicx}
\usepackage[pdftex]{xcolor}
\usepackage{subcaption}

\usepackage{tikz,pgfplots}
\usetikzlibrary{external,backgrounds,fit,calc,positioning,matrix,decorations.markings,arrows,arrows.meta,fadings,decorations.pathmorphing,shapes.geometric}
\usepgfplotslibrary{colormaps}

\definecolor{carBlue}{rgb}{0.2059,0.69412,0.83922}%
\definecolor{carRed}{rgb}{0.99608,0.29,0.23922}%
\definecolor{carGreen}{rgb}{0.63137,0.78431,0.25098}%
\definecolor{paperBlue}{rgb}{0.07059,0.69412,0.83922}%
\definecolor{paperGreen}{rgb}{0.63137,0.78431,0.25098}%
\definecolor{pathGreen}{rgb}{0.63137,0.78431,0.25098}%
\definecolor{pathRed}{rgb}{0.99608,0.44118,0.23922}%
\definecolor{paperRed}{rgb}{0.99608,0.74118,0.23922}%
\definecolor{paperOrange}{rgb}{0.96078,0.89804,0.14902}%

\def\straightX{{t}}
\def\straightY{{0}}
\def\rightX{{t}}
\def\rightY{{-((1)/(1+e^(-10*(t-0.45))))}}
\def\leftX{{t}}
\def\leftY{{((1)/(1+e^(-10*(t-0.45))))}}

\newlength{\carLength}
\newlength{\freeLength}
\newlength{\fwidth}
\def\definemappedcolor#1{%
    \pgfmathparse{#1*1000}
    \pgfplotscolormapdefinemappedcolor{\pgfmathresult}%
}%

\pgfplotsset{
    colormap={accBrColor}{
        color=(carGreen)
        color=(carRed)
    },
    colormap={brAccColor}{
    	color=(carRed)
        color=(carGreen)
    },
    colormap={Acc1Color}{
    	color=(carGreen!50)
        color=(carGreen!50)
    },
    colormap={Acc2Color}{
    	color=(carGreen!70)
        color=(carGreen!70)
    },
    colormap={Acc3Color}{
    	color=(carGreen)
        color=(carGreen)
    },
    colormap={Br1Color}{
    	color=(carRed!50)
        color=(carRed!50)
    },
    colormap={Br2Color}{
    	color=(carRed!70)
        color=(carRed!70)
    },
    colormap={Br3Color}{
    	color=(carRed)
        color=(carRed)
    },
    colormap={Norm1Color}{
    	color=(carBlue!50)
        color=(carBlue!50)
    },
    colormap={Norm2Color}{
    	color=(carBlue!70)
        color=(carBlue!70)
    },
    colormap={Norm3Color}{
    	color=(carBlue)
        color=(carBlue)
    },
}

\tikzset{
  set arrow inside/.code={\pgfqkeys{/tikz/arrow inside}{#1}},
  set arrow inside={end/.initial=>, opt/.initial=},
  /pgf/decoration/Mark/.style={
    mark/.expanded=at position #1 with
    {
      \noexpand\definemappedcolor{#1}%
      \noexpand\arrow[\pgfkeysvalueof{/tikz/arrow inside/opt}]{\pgfkeysvalueof{/tikz/arrow inside/end}}
    }
  },
  arrow inside/.style 2 args={
    set arrow inside={#1},
    postaction={
      decorate,decoration={
        markings,Mark/.list={#2}
      }
    }
  },
}

\DeclareMathAlphabet{\mathsfit}{\encodingdefault}{\sfdefault}{m}{sl}
\SetMathAlphabet{\mathsfit}{bold}{\encodingdefault}{\sfdefault}{bx}{sl}

\newcommand{\randScal}[1]{\mathsfit{#1}}

\title{\LARGE \bf
Unsupervised and Supervised Learning with the Random Forest Algorithm for Traffic Scenario Clustering and Classification
}

\author{Friedrich Kruber$^{1}$, Jonas Wurst$^{1}$, Eduardo S\'{a}nchez Morales$^{1}$,\\Samarjit Chakraborty$^{2}$, Michael Botsch$^{1}$
\thanks{$^{1}$Technische Hochschule Ingolstadt, Research Center CARISSMA, Esplanade 10, 85049 Ingolstadt, Germany,
\{firstname.lastname\}@thi.de\
$^{2}$Technical University of Munich, Real-Time Computer Systems, Arcisstra{\ss}e 21, 80333 Munich, Germany, \{firstname\}@tum.de}
}

\begin{document}

\maketitle
\thispagestyle{empty}
\pagestyle{empty}

\begin{abstract}
The goal of this paper is to provide a method, which is able to find categories of traffic scenarios auto\-matic\-ally. The architecture consists of three main components: A microscopic traffic simulation, a clustering technique and a classification technique for the operational phase. The developed simulation tool models each vehicle separately, while maintaining the dependencies between each other. The clustering approach consists of a modified unsupervised Random Forest algorithm to find a data adaptive similarity measure between all scenarios. As part of this, the path proximity, a novel technique to determine a similarity based on the Random Forest algorithm is presented. In the second part of the clustering, the similarities are used to define a set of clusters. In the third part, a Random Forest classifier is trained using the defined clusters for the operational phase. A thresholding technique is described to ensure a certain confidence level for the class assignment. The method is applied for highway scenarios. The results show that the proposed method is an excellent approach to automatically categorize traffic scenarios, which is particularly relevant for testing autonomous vehicle functionality.
\end{abstract}



\section{INTRODUCTION}
In recent work the scenario based approach is seen as one promising way to test automated vehicle functions \cite{ForschungsprojektPEGASUS.2017}. A kilometer based approach is not feasible for testing and validating functions for automated driving, which have to master all conditions within a given specification. That would take billions of kilometers to complete \cite{Shwartz2017, Zhao.2017, Maurer.2016b}. Functions, which are currently under development, show clearly the need for a relevance evaluation to limit the test expenditure \cite{Weitzel.2014}.
The test effort must be justified based on an analysis of the relevance of scenarios. Therefore, one must know which categories of scenarios appear on the road. Looking at scenarios microscopically, there is an infinite number of scenarios. A somewhat higher visual range shows that they nevertheless follow certain patterns and can be assigned to categories with a high degree of similarity. Testing representatives from each category ensures a broad scope, while minimizing the effort in the validation process. 

The main contribution of this paper is to describe a process of how to learn categories/clusters of traffic scenarios in an unsupervised way, given only the data. The methods presented in this work are separated into three main parts. 
The first element is a microscopic traffic simulation tool, which allows an accurate modeling of the vehicle motion and interaction between traffic participants. 

The second part is the extended (Modified) Unsupervised Random Forest (xMURF) used to determine a data adaptive similarity measure. The xMURF is an extension of the MURF method presented in \cite{Kruber.2018}. As a part of the xMURF, a novel similarity measure, based on the Random Forest (RF) is introduced, called path proximity. The difference to the normal RF proximity is that, given a datapoint the whole path through each tree of the RF is considered instead of just the terminal leaf. The similarities between the scenarios achieved this way are visualized as a reordered proximity matrix. One defines clusters/categories based on this visualization. 

As third and last part, the clusters are used to train a RF classifier. This way, new traffic scenarios can be assigned to a category. A special assignment rule is used, such that scenarios only get assigned to a cluster if they exceed a certain confidence threshold. The architecture is examined by using a data set of highway scenarios realized by the simulation tool. 

In summary, the work presents a method which enables one to investigate relations between traffic scenarios. It also enables one to define groups of scenarios, such that categories are created, which are used to classify new scenarios. Moreover, also other research fields can benefit from this work, since the presented xMURF is not limited to traffic scenario data.

The paper is organized as follows. Section \ref{relatedwork_section} depicts related work. Section \ref{architecture_section} describes the architecture of the proposed method. The key components of the simulation are described in Section \ref{data_section}. Next, a definition of a scenario and a description of the extracted features is given in Section \ref{data_section}. Section \ref{clustering_section} explains the xMURF method. The enhancements to its predecessor version \cite{Kruber.2018} are discussed as well. The classification of new scenarios is explained in Section \ref{classification_section}. In Section \ref{results_section} some demonstrational results of the xMURF method based on the simulation data are presented. Finally, this work is concluded in Section \ref{conclusions_section}.
\section{RELATED WORK}
\label{relatedwork_section}
This section depicts a selection of related work for both adjacent topics, the validation of automated vehicles and the clustering method.
In \cite{Koopman.2016}, a summary of the challenges of autonomous vehicles and proposals to deal with them is given.
A method to accelerate the validation process is proposed in \cite{Zhao.2017}. The research initiative PEGASUS \cite{ForschungsprojektPEGASUS.2017} examines the validation for automated vehicles from various aspects and aims to provide widely accepted standards.
In \cite{Wachenfeld.2016b, Wagner.2018} two measures for risk assessment are proposed to capture relevant traffic scenarios. Both publications describe methods to find trigger points to recognize relevant scenarios, while this work is focused on finding categories within the set of recognized scenarios. Trigger points are used in this work to define the length of a scenario as described in\ Section \ref{data_section}. 

The RF for classification and regression is presented in \cite{Breiman2001random}, while unsupervised usage is described in successing work \cite{Breiman02URF}.
In \cite{Cara.2015}, a classification for car-cyclist scenarios is presented. It is important to mention, that the classes are predefined in \cite{Cara.2015}.
An application of the unsupervised RF is presented in \cite{Shi2006}. In \cite{Liu2000}, a tree based clustering technique is shown, which was a starting point for the MURF algorithm \cite{Kruber.2018}. The related work on clustering differs by the techniques to generate noise data (see Section \ref{clustering_section}) and the way how the clusters are defined. Additionally, this paper proposes the path proximity to determine a similarity measure.
\section{ARCHITECTURE}
\label{architecture_section}
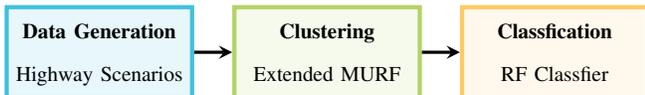
\begin{figure}[h]
\centering
\begin{footnotesize}
	\begin{tikzpicture}
\def\ownSpace{4ex}
\def\ownHeight{8ex}
\def\ownWidth{18ex}

\node(Generation)[fill=paperBlue!10,draw=paperBlue!80, ultra thick]{
\begin{minipage}[t][\ownHeight][t]{\ownWidth}\vspace{1ex}\centering
\textbf{Data Generation}\\[2ex] Highway Scenarios
\end{minipage}};

\node(Clustering)at($(Generation.east)+(\ownSpace,0)$)[anchor=west,fill=paperGreen!10,draw=paperGreen!80, ultra thick]{
\begin{minipage}[t][\ownHeight][t]{\ownWidth}\vspace{1ex}\centering
\textbf{Clustering}\\[2ex] Extended MURF
\end{minipage}};

\node(Analysis)at($(Clustering.east)+(\ownSpace,0)$)[anchor=west,fill=paperRed!10,draw=paperRed!80, ultra thick]{
\begin{minipage}[t][\ownHeight][t]{\ownWidth}\vspace{1ex}\centering
\textbf{Classfication}\\[2ex] RF Classfier
\end{minipage}};

\draw[-stealth,very thick] (Generation.east) -- (Clustering.west);

\draw[-stealth,very thick] (Clustering.east) -- (Analysis.west);

\end{tikzpicture}
	\end{footnotesize}
	\vspace{-0.3cm}
	\caption{The overall architecture}
	\label{fig:ToolChainComplete}
	\vspace{-0.0cm}
\end{figure}

In this section an overview of the overall architecture is given. The main steps are depicted in Fig. \ref{fig:ToolChainComplete}. 
To begin with, a data set has to be collected. Therefore, a simulation tool has been developed which generates traffic flow. The road is modeled as a highway with two or three lanes. The number of vehicles, as well as their behavior, is varied over many simulation loops. One element in varying the traffic situations is the induction of randomness to the simulation parameters. 

After each loop, an algorithm checks for interesting scenarios. Within the time line of a scenario, relevant features are extracted. The features are aligned into a vector $\bm{x} \in \mathbb{R}^Q$ and used as the input for the xMURF. This intermediate step compresses the relevant information from the time series data. The feature vectors from all scenarios are then processed by the xMURF algorithm. The learning process is determined by the selected features and the given data set $\mathcal{D}$. The proposed method makes usage of the path a datapoint is following in each tree in the RF. Comparing the paths between two datapoints delivers the necessary similarity measure. The output of the xMURF is a similarity matrix with similarity values between all pairs of scenarios. This matrix is then rearranged by means of hierarchical clustering and represented as a graphic, where each pixel is colored according to its similarity value. As similar scenarios are located nearby and share a similar color, clusters get visible. Based on the visualization, clusters can be identified and labeled. 

After assigning labels to each cluster, the class of each scenario can be stated. Taking again the input feature vectors with their new assigned label, one receives the data set $\mathcal{D}_\mathrm{c}$. The input data $\bm{x}$ is now assigned to a class $y$, so that 
\begin{equation}
\mathcal{D}_\mathrm{c}=\left\lbrace (\bm{x}_1,y_{1}),\dots,(\bm{x}_{M_\mathrm{c}}, y_{M_\mathrm{c}})\right\rbrace, 
\label{eq:Dclass}
\end{equation}
with ${M_\mathrm{c}}$ being the number of datapoints classified. Given $\mathcal{D}_\mathrm{c}$, now a supervised RF classifier is trained.  

In the operational phase, a new datapoint is only assigned to a class, if a class adaptive threshold is exceeded. The threshold is computed based on the out-of-bag (OOB) method \cite{Breiman96b}.
If a datapoint gets not assigned nor clustered, that indicates, that not enough examples are collected to form a category of similar traffic scenarios for the given setup.  
\section{DATA GENERATION}
\label{data_section}
This section describes the procedure of data generation to demonstrate the proposed method. It is divided into three parts: first the traffic simulation, followed by a definition for scenarios and finally a discussion about the chosen features.
\subsection{Traffic simulation}
\label{simulation_section}
In order to generate traffic flow data on a microscopic level, a simulation tool has been developed. In the previous work \cite{Kruber.2018}, SUMO \cite{SUMO_URL} was used and is replaced now for the following reasons. SUMO is optimized to simulate traffic flow for a large number of vehicles, while in this work the focus is put on more accurate vehicle models and controllers. Generally, SUMO generated traffic follows strictly accident free models, while for this work the vehicle properties are selected in such a fashion, that also highly critical constellations appear. This can rather be observed on roads than keeping guaranteed safe gaps. 

Some main ideas of the simulation are described in the following. First, all participants react to the other vehicles if their safeness is affected. Every vehicle is in charge of the gap ahead on its own lane. The main goal is an accident free drive, while striving for the maximum allowed velocity. Vehicles are allowed to overtake from both, the left and right handed side. Overtaking from the right side is permitted in some countries on highways in dense traffic situations  with velocities lower than $\unit[60]{km/h}$. However, this can also be observed at higher velocities. In the simulation, the target velocity is changed over time and chosen from a normal distribution  $v \sim \mathcal{N}(\unit[18]{m/s},\,\unit[4]{m/s}$). The mean value is selected lower than permitted speed limits, which is supposed to emphasize a dense traffic situation to further increase the chance of capturing relevant scenarios. 

The behavior setting includes the eagerness for higher accelerations and velocities, a risk profile which affects the accepted gaps and a politeness factor to cooperate with others during lane change maneuvers. The behavior setting changes during the simulation. As an example, a vehicle prevented from changing lanes caused by nearby vehicles on the target lane, might accept a smaller gap after some waiting time. Including such behavior models was inspired by \cite{SUMO_URL} and associated work. Depending on the vehicles settings, collisions might occur. 

One possibility to increase the variety of traffic scenarios generated, is the induction of randomness. Therefore, variables describing the vehicles abilities and its behavior, are shuffled on a random basis within defined boundaries. As an example, the maximum acceleration and jerk are limited based on real vehicle measurements. Next, a highway with $n_\mathrm{l}\in\{2,3\}$ number of lanes is generated. A random number of vehicles ($n_\mathrm{v}$) within the range of 
\begin{equation}
n_\mathrm{l} +1 \leq n_\mathrm{v} \leq n_\mathrm{l} \, n_\mathrm{vpl}, 
\end{equation}
where $n_\mathrm{vpl}$ is the maximum number of vehicles per lane, are placed on the road. 
As vehicles perform lane changes and overtake others, the unique index of each vehicle must be tracked for each time step $\mathrm{ts}$. These indices are stored in a three dimensional array $\bm{A} \in \mathbb{N}^{n_\mathrm{l} \times n_\mathrm{vpl} \times n_\mathrm{ts}}$. For modeling the lane change behavior several models are available, e.\,g. \cite{Treiber.2006}. The four motivations for lane-changing are described in \cite{Erdmann.2014}. Nevertheless, the motivation for a lane change in this simulation is mainly determined on a random basis. The purpose is not to model an ideal vehicle and traffic flow behavior. Lane changes are not always based on rational effectiveness. Additionally, wrong decisions can be made due to incomplete perception regarding both, humans and autonomous vehicles. As all vehicles aim towards an accident free travel, the lane changing vehicle checks for some empty space, while the accepted gap is set according to its risk and patience level. If the gap decreases below a threshold during the lane change, the vehicle aborts its maneuver.  

The longitudinal goals, which cause the acceleration at a certain time step, are determined by a leader vehicle. This behavior model is called the car-following model. Contrary to accident-free implementations, the distances in this simulation fall below an optimal safe gap to achieve a more natural approaching behavior and in order to generate critical scenarios. The maximum deceleration for a following vehicle, given optimal conditions with $a_{x} \approx \unit[-1]{g}$, is reached to eliminate the relative velocity at a distance close to zero. The leader vehicle acceleration is determined by the target velocity $v_{x,\mathrm{l}}$. Additionally, the inter-leader gap $d_\mathrm{il}$ in driving direction is bounded to $d_\mathrm{il,max}$ and overrules the target velocity, in order to keep all vehicles within a certain region and ensure dense traffic with a limited number of vehicles. The acceleration profiles are specified according to the Gompertz function 
\begin{align}
a_{x,\mathrm{f}}(d_\mathrm{fl}) &= a_\mathrm{m}e^{{-be}^{(-cd_\mathrm{fl})}},\\
a_{x,\mathrm{l}}(v_{x,\mathrm{l}}, d_\mathrm{il}) &= \left\lbrace \begin{array}{l l}
a_\mathrm{m}e^{{-be}^{(-c d_\mathrm{il})}} & d_\mathrm{il} > d_\mathrm{il,max} \\
a_\mathrm{m}e^{{-be}^{(-c v_{x,\mathrm{l}})}} &\text{else}\\
\end{array}\right.,
\end{align}
where the variable $a_{x}$ applies to the actual longitudinal acceleration. The three Gompertz function variables are $a_\mathrm{m}$, $b$ and $c$, where $a_\mathrm{m}$ applies to the maximum acceleration ability and $b$ and $c$ are shuffled. The index f indicates the following vehicle, l the leader vehicle, il the inter-leader vehicle relation and fl the follower leader vehicle relationship.
The Gompertz function implies a simplified model for the mass inertia at the beginning of the acceleration, followed by a linear increase and the limited torque or braking capacity towards the maximum acceleration or deceleration, respectively. Furthermore, a reaction time is added for each vehicle.
  
The lateral dynamics are governed by the combination of a one-track model (OT) and a controller that steers the vehicle to the middle of its current lane, or towards the middle of the neighboring lane in case of a lane change. First, the future pose of the vehicle is predicted for a velocity dependent look-ahead time using the current velocity and steering angle during the prediction horizon. An orientation error and distance error are computed by comparing the predicted pose with the desired one. These two quantities and a velocity dependent weight are used by a P-controller
to output the steering angle. Larger distance errors favour steering the vehicle towards the desired lane, and smaller distance errors favour correcting the vehicle orientation. Finally, the OT calculates the updated state variables, which allows modeling the driving dynamics approriatly up to a lateral acceleration of $a_{y} \approx \unit[0.4]{g}$ \cite{Ammon.1997}. This is sufficient for the simulated lateral maneuvers.
\subsection{Scenario Definition}
\label{scenario_subsection}
A traffic scenario can be defined in different ways. A literature review and a proposal of a scenario definition in the sense of context modeling is given in \cite{Ulbrich.2015}. Still, up to now there is no widely accepted, unified definition available. For this work, the general understanding of a scenario is a sequence of connected events. It implies the entities and their state transitions over time. 
The entities with dynamically changing states are the ego-vehicle $ego$ and the surrounding vehicles $tg$ in this work. The $tg$ are evaluated based on the relevance for the $ego$. All vehicles have their own dynamic model, the state is updated each time step. The scenery, i.\,e. the static objects, is limited to the highway road with its number of lanes and its meta data, which is here the permitted velocity. Hence, the environment model of the scenario is described in an object-based form. All relevant objects in the vicinity of the $ego$ are tracked and evaluated over time. An alternative form would be the grid-based form, which discretizes the static environment into cells of equal size. The existence of objects is indicated by the occupancy of the cells in the stationary raster map \cite{Maurer.2016b}.    

The starting point of a scenario for this work is set if a criticality measure is exceeded and ends whenever the value falls below the threshold again. Hence, the scenario contains an arbitrary number of entities and maneuvers in an arbitrary period of time. 

Plenty of options to choose a criticality measure can be utilized, for example \cite{Wachenfeld.2016b, Wagner.2018}. In \cite{Kitajima.2009} several measures are compared, including the common Time-To-Collision (TTC) and Time-Headway (THW). The THW is calculated by the relative distance over the $ego$ velocity. The TTC is introduced as the relative distance divided by the relative velocity, which is difficult to estimate for humans. The paper \cite{Kondoh.2008} analyzes THW and TTC more detailed: The THW determines the car-following strategies of drivers and the magnitude of the THW influences their stress level.
Because of its intuitiveness and its reflection of the stress level for drivers or passengers in automated vehicles, the THW is chosen here.
 
In this work, the recording of a scenario begins far beyond criticality. It is rather an undercutting a comfort zone, which is set to $\mathrm{THW} \leq \unit[1.0]{s}$. As a constraint, the scenario is withdrawn if $\mathrm{THW_{min}} > \unit[0.8]{s}$ to retrieve scenarios with some inter-vehicle dynamics, where $\mathrm{THW_{min}}$ corresponds to the minimum THW in the scenario. The goal is to record the evolving situation from a normal state to its maximum criticality and the recording is continued until the scenario resolves and reaches either a comfortable state again or ends in a collision. In order to be able to better judge the numbers, following the administrative rule of thumb for German highways would yield a $\mathrm{THW} \approx \unit[1.8]{s}$. Approaching the leader vehicle closer than $\mathrm{THW} \leq \unit[0.9]{s}$ is punishable because of violating a safe distance \cite{Filzek.2002}.
\subsection{Feature generation}
The features are derived from both, time and space considerations of the scenario. They all share some relevance to the $ego$ and can be recorded by state of the art vehicles. It is assumed, that the data source is exclusively the $ego$ sensors and no presence of V2X communication. The goal is to find a set of features, which gives an appropriate balance to capture the environment in 360 degrees over time, while still being compact for fast computation. Some features are discussed in this section, the complete set is available under \cite{KruberGithub.2019}.

As mentioned before, the existence of a scenario is defined by its criticality magnitude, which results in arbitrary time spans. Therefore, the features which are related to the time dimension, are set up on three time instances: At the beginning and the end of a scenario and at the critical changepoint, which is here defined by $\mathrm{THW_{min}}$. Other features observe changes over time. As an example, the actual distance to the leader vehicle is compared to the desired distance over time by utilizing the dynamic time warping function, which compares the shapes of two curves. A high discrepancy between both curves indicates high dynamics between both vehicles and thus criticality.
 
\begin{figure}
\par\smallskip
\centering
  \input{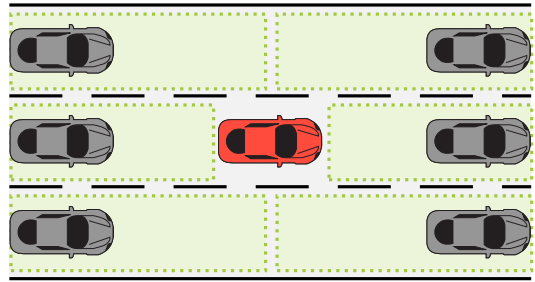}
	\caption{The movement of the ego vehicle (red) is determined by the six surrounding vehicles. Therefore the surrounding space is divided into six regions. The size of the regions is adapted to the velocity.}
	\vspace{-0.2cm}
	 \label{fig:vehicles}
	 \vspace{-0.0cm}
\end{figure}

The area around the $ego$ is divided into six zones. It is assumed, that the maximum number of vehicles, which influence the $ego$ directly and vice versa, is limited to six. This proposition is depicted in Fig. \ref{fig:vehicles}, where the red vehicle represents the $ego$. The six surrounding vehicles determine its moving options in longitudinal and lateral direction. As a consequence, the relative distance to all of those vehicles is important and added to the feature set. The desired relative distances act as simplified forms of safety zones, which are schemed by the green boxes in Fig. \ref{fig:vehicles}. The safety zones determine the $ego$s options to plan its trajectory and prevents the vehicles from contact. The main attention while defining features, however, is focused on the interaction with the leader vehicle.

Other features are defined according to the semantics of the road, which are in a simplified highway case the number of lanes. Features indicating a lane change of the $ego$, as well as changes of a $tg$, which result in a cut-in maneuver from the $ego$ perspective are captured. Also the lane index of the $ego$ is taken into account: it emphasizes, if the $ego$ had the chance to escape the threatening situation by performing a lane change onto the left or right side.
In total a set of 47 features are selected to be fed in the clustering process. The description of all features, along with the recorded features for all generated scenarios, are given in \cite{KruberGithub.2019}.
\section{CLUSTERING}
\label{clustering_section}
The aim of this work is to find scenario categories in an unlabeled data set. For this purpose, the clusters are defined manually by examining a sorted proximity matrix $\bm{P}_\mathrm{o}$ visually. The unsorted proximity matrix  $\bm{P}$ itself is generated based on the xMURF algorithm, which is presented in this section.
\subsection{The Extended Unsupervised Random Forest}
In the previous work \cite{Kruber.2018}, a modification of the RF algorithm \cite{Breiman2001random} for unsupervised usage is shown, called MURF. The MURF algorithm is based on the Unsupervised RF (URF) \cite{Breiman02URF,Shi2006}. The aim of the URF is to provide a data adaptive similarity measure in terms of the RF proximity. In \cite{Kruber.2018} also the graphical interpretation of the resorted proximity matrix is explained, where the resorting is realized through hierarchical clustering. In this paper, some extensions to this technique are shown. 

The URF is used to generate a data adaptive similarity measure. In order to perform unsupervised learning with a RF, a noise data set $\mathcal{S}$ is generated based upon the given data set $\mathcal{D}$. The RF is trained to solve the classification task, where it has to be distinguished between the given data set $\mathcal{D}$, labeled as one class, and the generated noise data set $\mathcal{S}$, labeled as a different class. This way, the trees have to fit their leaves to the given data set $\mathcal{D}$. Extracting the RF proximity for the given data set leads to a data adaptive similarity measure. In the MURF algorithm proposed in \cite{Kruber.2018}, the noise data is not generated, instead it is assumed in each split. The used distribution to calculate the necessary number of noise datapoints is based on an uniform distribution. 

On the one hand side, the MURF algorithm enables one to adjust the granularity of the proximity measure by adjusting the impurity level, which is used as the pruning criterion. On the other hand side, the results are very sensitive to the choice of this hyperparameter, leading to a tedious tuning process. Since the RF proximity just examines the terminal leaves, the pruning has a large influence on the resulting similarity measure.

In this paper two important extensions of the MURF algorithm are introduced. Both are summarized in the algorithm called xMURF. First, the pruning impurity parameter is removed by extracting the similarity out of the datapoints' paths through the trees instead of only taking the terminal leaves into account. Therefore, the path proximity is introduced in this paper. In xMURF the complete information regarding the similarity provided by the RF is captured, which makes the process robust. Second, the dependency on the uniform distribution in MURF, used to calculate the splits, is optimized. In the xMURF algorithm the used distribution at each split is randomly chosen from a set of predefined probability density functions.

The proximity matrix $\bm{P}$ generated by the xMURF algorithm is reordered by hierarchical clustering, following the explanations in \cite{Kruber.2018, Behrisch2016}. For this purpose, one has to extract the leaf order of the dendrogram, which represents the hierarchical clustering. Then, the matrix $\bm{P}$ is resorted corresponding to the leaf order of the dendrogram, such that the similarity between the neighbors gets higher. If necessary, an optimal leaf ordering algorithm can be applied to refine the sorting further.

The reordered matrix $\bm{P}_\mathrm{o}$ is then examined graphically, in order to identify clusters as well as inter-cluster relationships. Some general rules for the interpretation are also explained in \cite{Kruber.2018} and more detailed in \cite{Behrisch2016}. Clusters will appear as bright squares (high similarity) along the diagonal of the proximity matrix. An example is depicted in Fig. \ref{fig:prox_Matrix}.
\subsection{Path Proximity}
This work introduces a novel proximity measure for the RF algorithm. The proximity measure takes into account the full paths of the datapoints through the trees instead of just using the terminal leaves. In \cite{Englund.2012} an alternative proximity measure for the RF was introduced, where the distance between two terminal leaves is examined.

Let a RF consist of $B$ trees $T$, where the $b$th tree $T_b$ is constructed based on the bagged data set $\mathcal{D}_b$. Then a tree $T_b$ consists of $N_b$ nodes $t_{n,b}$. A path of a datapoint through a tree can be defined by a set including all nodes the datapoint passed. This leads to the path formulation as
\begin{equation}
\mathcal{T}_{i,b} = \left\lbrace t_{1,b},t_{n_{i_2},b},\dots,t_{N_i,b}\right\rbrace,
\end{equation}
where $i$ is an index, representing the $i$th data point $\bm{x}_i$. The node $t_{1,b}$ is the root node of the $b$th tree $T_b$ and hence the first node on the path of the $i$th datapoint. The node $t_{n_{i_2},b}$ is the second node on the path, where $n_{i_2}$ represents the node number $n$ the datapoint has passed. The last node on the path of the $i$th data point in the $b$th tree is $t_{N_i,b}$.

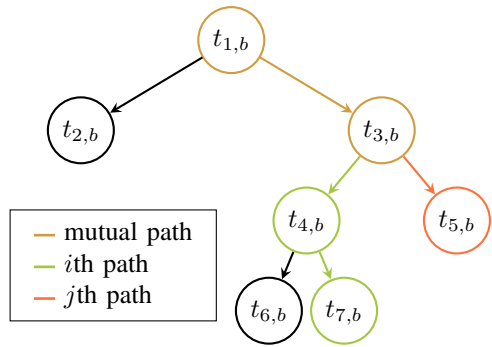
\begin{figure}
\par\medskip
\centering
	\begin{tikzpicture}[
	every node/.style = {circle,draw,minimum size=4ex,text depth=.3ex},
	level distance = 12mm,
	level 1/.style = {sibling distance=40mm},
	level 2/.style = {sibling distance=20mm},
	level 3/.style = {sibling distance=10mm},
]
\node[pathRed!50!pathGreen] (root) {\textcolor{black}{$t_{1,b}$}} [-stealth,thick,pathRed!50!pathGreen]
	child[black] {
		node(left) {\textcolor{black}{$t_{2,b}$}}		
		child[missing] {
			node {}
			child[missing] 	{node {}}
			child[missing] 	{node {}}
		}	
		child[missing] {
			node {}
			child[missing] {node {}}
			child[missing] {node {}}
		}
	}
	child{
		node {\textcolor{black}{$t_{3,b}$}} 
		child[pathGreen] {
			node {\textcolor{black}{$t_{4,b}$}}
			child[black] {node {\textcolor{black}{$t_{6,b}$}}}
			child {node {\textcolor{black}{$t_{7,b}$}}}
		}
		child[pathRed] {
			node {\textcolor{black}{$t_{5,b}$}}
			child[missing]	 	{node {}}
			child[missing] 	{node {}}
		}
	};

\node(legend)[at={($(left.south)+(0,-0.6cm)$)},anchor=120,rectangle,draw,inner sep=3pt,text depth={}]{\begin{tabular}{l}
\textcolor{pathRed!50!pathGreen}{\rule[0.3ex]{8pt}{1pt}} mutual path\\
\textcolor{pathGreen}{\rule[0.3ex]{8pt}{1pt}} $i$th path\\
\textcolor{pathRed}{\rule[0.3ex]{8pt}{1pt}} $j$th path
\end{tabular}
};
\end{tikzpicture}
	\caption{Path proximity example}
	\vspace{-0.3cm}
	\label{fig:pathProximity}
	\vspace{-0.0cm}
\end{figure}

In order to compare the paths of two datapoints $i$ and $j$ through the $b$th tree, the corresponding sets $\mathcal{T}_{i,b}$ and $\mathcal{T}_{j,b}$ need to be compared. In this work, the Jaccard Index \cite{Jaccard.1912}
\begin{align}
P_{ij}\left(b\right) &= \frac{\vert\mathcal{T}_{i,b}\cap\mathcal{T}_{j,b}\vert}{\vert\mathcal{T}_{i,b}\cup\mathcal{T}_{j,b}\vert}\\
&=\frac{\vert\mathcal{T}_{i,b}\cap\mathcal{T}_{j,b}\vert}{\vert\mathcal{T}_{i,b}\vert+\vert\mathcal{T}_{j,b}\vert-\vert\mathcal{T}_{i,b}\cap\mathcal{T}_{j,b}\vert}\label{eq:pathProxSingleTree}
\end{align}
is used for this purpose. It holds that $P_{ij}(b)\in(0,1]$, since at least the root node is present in both sets. This way, a similarity measure, given the two datapoints $i$ and $j$, based on the $b$th tree is defined. In Fig. \ref{fig:pathProximity} an example tree with two arbitrary paths is shown. The mutual path of both datapoints is marked in ocher, where the single paths are shown in green and red. Interpreting Eq. (\ref{eq:pathProxSingleTree}) based on Fig. \ref{fig:pathProximity} leads to $\vert\mathcal{T}_{i,b}\cap\mathcal{T}_{j,b}\vert$ being the length of the mutual path and $\vert\mathcal{T}_{i,b}\vert$ being the length of the $i$th path ($i$th datapoint) starting from the root node, $\vert\mathcal{T}_{j,b}\vert$ respectively. For the given example, the corresponding Jaccard index would be $\nicefrac{2}{5}$. Or in other words, the $i$th and $j$th datapoints have a similarity of $0.4$. A value of $1$ indicates, that both datapoints are identical or very similar according to the given tree.

By averaging over all $B$ values $P_{ij}\left(b\right)$ in a forest, one achieves the RF path proximity
\begin{equation}
P_{ij}=\frac{1}{B}\sum\limits_{b=1}^B \frac{\vert\mathcal{T}_{i,b}\cap\mathcal{T}_{j,b}\vert}{\vert\mathcal{T}_{i,b}\vert+\vert\mathcal{T}_{j,b}\vert-\vert\mathcal{T}_{i,b}\cap\mathcal{T}_{j,b}\vert}.
\end{equation}
If two datapoints have the same paths in all trees the proximity will be one. Contrary, if they only share the root nodes, the proximity will be very small tending towards zero, the longer both paths are. The RF path proximity enables one to cover more than just the leaf information of the forest within one scalar value.

As already mentioned, the path proximity removes the highly sensible impurity pruning parameter from the MURF algorithm. Fully grown, instead of pruned trees can be used, since they lead to a more meaningful similarity measure.
\subsection{Ensemble Noise}
In this work, a random selection of noise distribution at each split is performed, called ensemble noise. During the construction of the xMURF, the used distributions are chosen randomly from a collection of predefined distributions.

The estimated Gini impurity for the node $t$ in an arbitrary tree is defined as
\begin{equation}
r(t) = \sum\limits_{c=1}^C \frac{M_c(t)}{M(t)}\left(1-\frac{M_c(t)}{M(t)}\right),
\end{equation}
where $C$ is the number of classes (2 for the xMURF) and $M_c(t)$ the number of datapoints in node $t$, which belong to class $c$. Based on the Gini impurity, the Gini gain $\Delta R$ resulting by splitting $t$ into the child nodes $t_\mathrm{L}$ (left) and $t_\mathrm{R}$ (right) is defined by
\begin{equation}
\Delta R(t,t_\mathrm{L},t_\mathrm{R}) = r(t)-\frac{M(t_\mathrm{L})}{M(t)}r(t_\mathrm{L})-\frac{M(t_\mathrm{R})}{M(t)}r(t_\mathrm{R}).
\end{equation}
The optimal split is given if $\Delta R(t,t_\mathrm{L},t_\mathrm{R})$ is maximal. Hence, the number of datapoints of each class in each node ($t$, $t_\mathrm{L}$ and $t_\mathrm{R}$) is required. The number of original datapoints $M_{\mathcal{D}_b}\left(t_{n,b}\right)$ of the bagged data set $\mathcal{D}_b$ belonging to the $b$th tree in the $n$th node $t_{n,b}$ of this tree, as well as in the possible child nodes can simply be counted. The number of noise datapoints in the same node needs to be estimated for a given split value $\tau_{\tilde{q}}$. Let $z_{\tilde{q}}$ be the standardized value of $\tau_{\tilde{q}}$ (see Eq. (\ref{eq:tauz})). Then, the number of noise datapoints for the left and right child nodes of a given node $t_{n,b}$ is calculated as
\begin{align}
M_{\mathcal{S},\mathrm{l}_{n,b}}\left(z_{\tilde{q}}\right) &= M_{\mathcal{D}_b}\left(t_{n,b}\right) \mathrm{P}\left(\randScal{z}_{\tilde{q}}\leq z_{\tilde{q}}\right)\text{ and}\\
M_{\mathcal{S},\mathrm{r}_{n,b}}\left(z_{\tilde{q}}\right) &= M_{\mathcal{D}_b}\left(t_{n,b}\right)-M_{\mathcal{S},\mathrm{l}_{n,b}}\left(z_{\tilde{q}}\right),
\end{align}
where $\tilde{q}$ stands for the $\tilde{q}$th dimension of the vector whose features are chosen randomly in each node when constructing the trees in a RF. The number of features that are used at each split of each tree in the RF is $\tilde{Q}\in\mathbb{N}$, with $\tilde{Q}=\lfloor\sqrt{Q}\rfloor$. The values $M_{\mathcal{S},\mathrm{l}_{n,b}}$ and $M_{\mathcal{S},\mathrm{r}_{n,b}}$ denote the number of corresponding noise points in the left and right child note of $t_{n,b}$, given that the split $\tau_{\tilde{q}}$ is chosen, since $z_{\tilde{q}}$ is the standardized value of $\tau_{\tilde{q}}$. $ \mathrm{P}\left(\randScal{z}_{\tilde{q}}\leq z_{\tilde{q}}\right)$ is the value of the cumulative density function at the standardized threshold $ z_{\tilde{q}}$. The standardized threshold $ z_{\tilde{q}}$ in the $\tilde{q}$th RF dimension is determined with
\begin{align}
z_{\tilde{q}} &=\frac{\tau_{\tilde{q}}-\mu_{\tilde{q}}}{\sigma_{\tilde{q}}},\label{eq:tauz}\\
\mu_{\tilde{q}} &=\frac{\mathrm{max}\left\lbrace \mathcal{X}_{t_{n,b}}\right\rbrace_{\tilde{q}}+\mathrm{min}\left\lbrace \mathcal{X}_{t_{n,b}}\right\rbrace_{\tilde{q}}}{2},\\
\sigma_{\tilde{q}} &=\frac{\mathrm{max}\left\lbrace \mathcal{X}_{t_{n,b}}\right\rbrace_{\tilde{q}}-\mathrm{min}\left\lbrace\mathcal{X}_{t_{n,b}}\right\rbrace_{\tilde{q}}}{6},
\end{align}
where $\mathrm{max}\left\lbrace \mathcal{X}_{t_{n,b}}\right\rbrace_{\tilde{q}}$ and $\mathrm{min}\left\lbrace \mathcal{X}_{t_{n,b}}\right\rbrace_{\tilde{q}}$ yield the maximum or minimum value of the subspace $\mathcal{X}_{t_{n,b}}$ in the dimension specified by $\tilde{q}$. The interval of a node covers $\pm 3\,\sigma_{\tilde{q}}$.

The first distribution used is the uniform distribution, where its cdf is given by
\begin{equation}
\mathrm{P}_\mathrm{u}\left(\randScal{z}_{\tilde{q}}\leq z_{\tilde{q}}\right)=\frac{1}{6}z_{\tilde{q}}+\frac{1}{2}.
\end{equation}
The standard normal distribution is the second used distribution and is approximated through \cite{Waissi.1996}
\begin{equation}
\mathrm{P}_\mathrm{n}\left(\randScal{z}_{\tilde{q}}\leq z_{\tilde{q}}\right)=\frac{1}{1+e^{-\sqrt{\pi}\left(\beta_1 z_{\tilde{q}}^5+\beta_2 z_{\tilde{q}}^3+ \beta_3 z_{\tilde{q}}\right)}},
\end{equation}
where $\beta_1=-0.0004406$, $\beta_2=0.04181198$ and $\beta_3=0.9$ holds.
Third, a bimodal distribution is used, which is build as the sum of two shifted standard normal distributions $\mathrm{P}_\mathrm{n}$ with
\begin{equation}
\mathrm{P}_\mathrm{b}\left(\randScal{z}_{\tilde{q}}\leq z_{\tilde{q}}\right)=\mathrm{P}_\mathrm{n}\left(\randScal{z}_{\tilde{q}}\!-\!3\leq z_{\tilde{q}}\!-\!3\right)+\mathrm{P}_\mathrm{n}\left(\randScal{z}_{\tilde{q}}\!+\!3\leq z_{\tilde{q}}\!+\!3\right).
\end{equation}

The randomly selected noise distributions at each split relax the dependency of the proximity measure to one specific distribution. Clearly, it is also possible to add other distributions to the assortment.
\section{CLASSIFICATION}
\label{classification_section}
This section describes, how the clustering results can be used for the operational phase, where new datapoints are compared to the known clusters.
Once the xMURF is trained with a data set $\mathcal{D}$ and the clusters are selected, they get assigned with a label resulting in a data set $\mathcal{D}_\mathrm{c}$ from Eq. (\ref{eq:Dclass}).
Then a RF classifier is trained with $\mathcal{D}_\mathrm{c}$.

A class adaptive threshold is calculated as follows. The OOB method is used to determine $\kappa_i$, which is the proportion of the votes for the correct class for the $i$th datapoint. The class adaptive threshold $\bar{\kappa}_c$ is computed as the average value of all $\kappa_i$ belonging to the class $c$. 

Next, the trained RF is used to predict the class assignment for a datapoint $\bm{x}_{i,\mathrm{new}}$ from a new data set $\mathcal{D}_\mathrm{new}$. The proportion of the votes of all trees for the winning class $c$ is compared to the class adaptive threshold $\bar{\kappa}_{c}$. If the value is below the threshold $\bar{\kappa}_{c}$, the assignment is withdrawn.

The threshold $\bar{\kappa}_{c}$ is dependent to the choice of clusters. Using an adjustable ratio of $\bar{\kappa}_c$, enables one to select an appropriate class assignment threshold, such that it fits to the given requirements. The ratio can be used to loosen or increase the strictness for the decision of assigning or withdrawing new datapoints to a class. 

Setting the threshold equally for all clusters, the assignment rate would be favored towards clearly separable clusters.
\section{RESULTS}
\label{results_section}
\begin{figure}
\par\medskip
\centering
\input{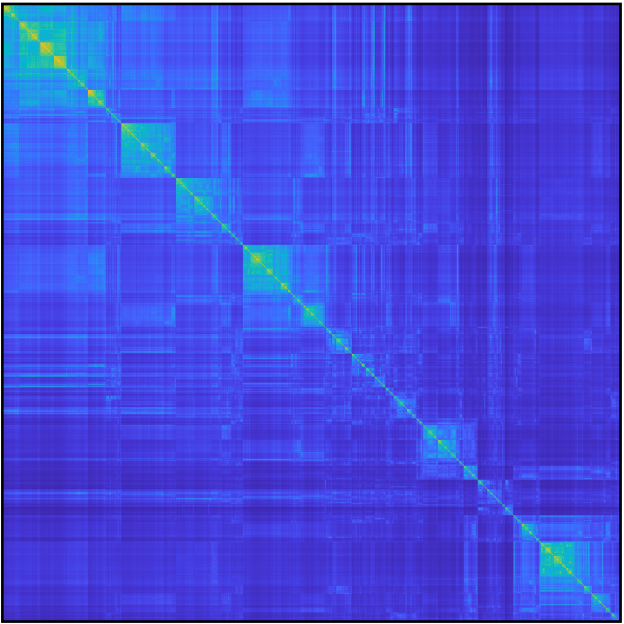}
	\caption{The resulting proximity matrix of data set $\mathcal{D}_{1}$, where yellow pixels indicate a high similarity. Clusters are represented as bright squares along the diagonal. Inter-cluster similarities are depicted as bright areas outside the clusters.}
	\vspace{-0.3cm}
	\label{fig:prox_Matrix}
	\vspace{-0.0cm}
\end{figure}

This section discusses results of the proposed architecture using an example data set $\mathcal{D}_{1}$ with $ M = 5131$ generated scenarios. The scenarios can be divided into three main types: The $ego$ approaching its leader vehicle while following its lane (type $\mathrm{A}$), the $ego$ performing a lane change and violating the safety zone after the lane change (type $\mathrm{L}$) and a cut-in maneuver performed by a $tg$ causing the violation (type $\mathrm{C}$). A combination of type B and C also appears in some clusters. 
The xMURF was trained with $B = 300$ trees. The symmetric proximity matrix $\bm{P}_\mathrm{o}$ of number $M \times M$ is depicted in Fig. \ref{fig:prox_Matrix}, where yellow pixels indicate a high similarity, dark blue pixels a low similarity. Each pixel contains the value of the similarity measure of two compared scenarios.   
A zoom-in view on the graphic reveals, that bigger dominant cluster contain smaller clusters. For demonstration purposes in total 100 clusters were manually chosen. The clustered scenarios can be downloaded and plotted in Matlab \cite{KruberGithub.2019}. The question how many clusters to choose is a problem specific one and therefore has to be adjusted to the requirements. The clusters can be selected by the following criteria: the minimum number of datapoints within a cluster, the minimum similarity value within that cluster, the homogeneity within that cluster and the deviation of the inter-cluster similarity in the local region around that cluster. The inter-cluster similarity is indicated by the brightness of the pixels outside the clusters. The next sub-section will give examples on some clusters.
\subsection{Scenario cluster examples}
\begin{figure}
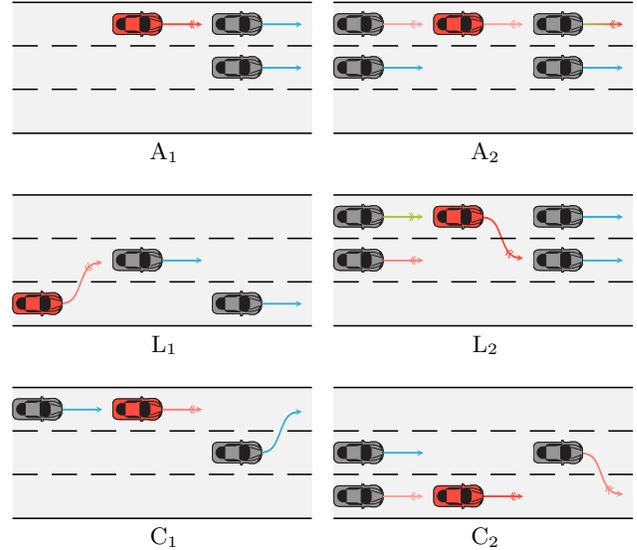

\par\medskip
\captionsetup[subfigure]{labelformat=empty}
	\center
	\vspace{-0.24\carLength}
	\begin{subfigure}[c]{0.48\columnwidth}
 		\resizebox{\columnwidth}{!}{\input{Approach1.tex}}
 		\vspace{-0.9cm}
 		\caption{$\mathrm{A_{1}}$}
	\end{subfigure}
	\begin{subfigure}[c]{0.48\columnwidth}
 		\resizebox{\columnwidth}{!}{\input{Approach2.tex}}
 		\vspace{-0.9cm}
 		\caption{$\mathrm{A_{2}}$}
	\end{subfigure}
	\begin{subfigure}[c]{0.48\columnwidth}
		\resizebox{\columnwidth}{!}{\input{LaneChange2.tex}}
		\vspace{-0.9cm}
		\caption{$\mathrm{L_{1}}$}
	\end{subfigure}
	\begin{subfigure}[c]{0.48\columnwidth}
 		\resizebox{\columnwidth}{!}{\input{LaneChange1.tex}}
 		\vspace{-0.9cm}
 		\caption{$\mathrm{L_{2}}$}
	\end{subfigure}
	\begin{subfigure}[c]{0.48\columnwidth}
		\resizebox{\columnwidth}{!}{\input{CutIn2.tex}}
		\vspace{-0.9cm}
		\caption{$\mathrm{C_{1}}$}
	\end{subfigure}
	\begin{subfigure}[c]{0.48\columnwidth}
 		\resizebox{\columnwidth}{!}{\input{CutIn1.tex}}
 		\vspace{-0.9cm}
 		\caption{$\mathrm{C_{2}}$}
	\end{subfigure}
		\caption{Each row depicts two examples for each scenario type ($\mathrm{A}$, $\mathrm{L}$, $\mathrm{C}$). The red vehicle depicts the $ego$, all $tg$ are colored gray. The arrow indicates the driving direction. The color of the arrow visualizes the acceleration, where green stands for positive, red for negative acceleration and blue for steady drive. The two arrows in the middle of the lines additionally indicate the accelerations, positive as well as negative.}
	\label{fig:ScenarioExamples}
	\vspace{-0.2cm}
\end{figure}

This section discusses examples of clusters, that result by applying the presented method. The example scenarios are depicted in Fig. \ref{fig:ScenarioExamples}. The figure depicts in total 6 scenarios, each one is a representative of its cluster.
Note, that a $tg$ (gray) is only shown in case it is located on the neighboring lane of the $ego$ (red). The first row depicts two scenarios of type approach (type A). In Fig. \ref{fig:ScenarioExamples}-$\mathrm{A_{1}}$, the $ego$ approaches the leader vehicle with a high relative velocity and brakes hard. This is a typical example for the A type scenarios, where the relative velocity and distance were estimated inappropriately by the $ego$. Fig. \ref{fig:ScenarioExamples}-$\mathrm{A_{2}}$ depicts a similar scenario, yet the consequences of such a scenario differs: first, the rear $tg$ on the same lane as the $ego$ does get affected and has to brake, which might emerge to a critical situation in a second phase. Second, the $ego$ does not have the chance to change the lane without risking a subsequent critical situation, as it would be enclosed by two other $tg$. That could cause a scenario as depicted in the below sub-figure $\mathrm{L_{2}}$.
The second row depicts two lane change scenarios. In Fig. \ref{fig:ScenarioExamples}-$\mathrm{L_{1}}$ the decision of $ego$ does not affect its environment. Contrary, in Fig. \ref{fig:ScenarioExamples}-$\mathrm{L_{2}}$ it causes the rear vehicle on the middle lane to brake, while the rear vehicle on the left lane starts accelerating due to the now given empty space, which closes the neighboring space on the left lane for the ego vehicle. 
The last row shows two cut-in maneuvers. Fig. \ref{fig:ScenarioExamples}-$\mathrm{C_{1}}$ depicts a typical cut-in maneuver. Due to a relatively high distance to the new leader (the cut-in performing $tg$ in the very front), the $ego$ is only affected slightly and brakes decently, illustrated by a slightly transparent red arrow. In Fig. \ref{fig:ScenarioExamples}-$\mathrm{C_{2}}$ the cut-in performing $tg$ brakes, forcing the $ego$ to perform a hard braking maneuver, which in consequence forces the rear vehicle to react as well. A risk mitigating lane change maneuver of the $ego$ is limited to the left side and would influence the rear vehicle on the middle lane.

Summarizing the above, all three scenarios in the left column show typical scenarios chosen out of the resulting automatically generated clusters for demonstration purposes of each type. The risk level during those scenarios is rather low and the $ego$ could theoretically perform a lane change to further mitigate the risk level it is exposed to. Also, its behavior does not directly influence its surrounding vehicles. On the right side, the same scenario types are depicted. Still, the $ego$ has less options for risk mitigating actions and its decisions influence the surrounding vehicles. These examples demonstrate, that dividing scenarios into categories has to include all relevant entities and scenery features, which results in a large variety of categories. Obviously, discussing the scenarios on this abstracted level, the risk and the influence of the $ego$ decision is correlated to the traffic density, which is consistent with the observations on real highways. The important aspect is, that these structures can be identified in an unsupervised, data adaptive way by means of a very compact scenario description. The data set is provided to the xMURF algorithm without any other additional information or pre-defined rules. The set of clusters available at \cite{KruberGithub.2019} show many more constellations. The clusters differ in the number of vehicles, but also in other aspects as the velocity, accelerations and the general vehicle behavior described in Section~\ref{simulation_section}.        
\subsection{Scenario classification}
To demonstrate how the selected clusters can be used to classify new scenarios, a new data set $\mathcal{D}_{2}$ is generated and processed as described in Section \ref{classification_section}.
In Table \ref{tb:Tabelle} the assignment rates for different ratios of $\bar{\kappa}_{c}$, given the data set $\mathcal{D}_{2}$, are depicted. 
\begin{table}[h]
\centering
\begin{tabular}{ | c | c |}
  \hline
  Ratio of $\bar{\kappa}_{c}$ & Assigned Scenarios\\
  \hline 
  1.00 & 34\%\\
  \hline
  0.75 & 55\%\\
  \hline
    0.50 & 81\%\\
  \hline
    0.25 & 98\%\\
    \hline
\end{tabular}
\caption{Assigned scenarios of validation the data set $\mathcal{D}_{2}$}
\label{tb:Tabelle}
\end{table}
Reducing the ratio leads to more datapoints being assigned, since the class membership criterion is less strict. The ratio of $\bar{\kappa}_{c}$ can be considered as an hyperparameter to adjust the required confidence measure of belonging to the class.

The classified scenarios of $\mathcal{D}_{2}$ can be compared with those from $\mathcal{D}_{1}$ by using the provided plot function, or by analyzing the data sets. The data sets are available at \cite{KruberGithub.2019}.
\section{CONCLUSIONS}
\label{conclusions_section}
The presented paper proposes an architecture to cluster traffic scenarios in an unsupervised way. The xMURF reveals the structure of the data and enables one to find similar scenarios and groups them in clusters. Compared to the previously presented work, this version is improved by introducing a path proximity and the ensemble noise method. 
The method was examined with traffic simulation data. The main principles of the simulation and generation of the data is described in this paper. Additionally, the scenarios are available under \cite{KruberGithub.2019}. Finally, an approach to classify new scenarios based on a class dependent threshold, which takes the initial clustering process into account, is presented. 


The proposed architecture can also be applied for other purposes, where one is interested in finding patterns of complex data sets and the output is not known beforehand.

\addtolength{\textheight}{-12cm}   



\section*{ACKNOWLEDGMENT}
The authors acknowledge the financial support by the Federal Ministry of Education and Research of
Germany (BMBF) in the framework of FH-Impuls (project number 03FH7I02IA). The authors thank the AUDI AG department for Testing Total Vehicle for supporting this work.

\bibliographystyle{IEEEtran}
\bibliography{ref}
\end{document}